\documentclass[10 pt, journal]{IEEEtran}

\usepackage{cite}
\usepackage{amsmath}
\usepackage{mathptmx} % assumes new font selection scheme installed
\usepackage{amssymb}  % assumes amsmath package installed
\usepackage{amsthm}
\usepackage{algorithmic}
\usepackage{array}
\usepackage{url}
\usepackage{graphics} % for pdf, bitmapped graphics files
\usepackage{epsfig} % for postscript graphics files
\usepackage{enumerate}
\usepackage{hyperref}
\usepackage{booktabs}
\usepackage[bottom]{footmisc}
\hypersetup{
	colorlinks=true, % false: boxed links; true: colored links
    linkcolor=blue,  % color of internal links (change box color with linkbordercolor)
    citecolor=blue,  % color of links to bibliography
    filecolor=blue,  % color of file links
    urlcolor=blue    % color of external links
}

\title{Comparison of Classical and Nonlinear Models for Short-Term Electricity Price Prediction}

\author{Authors}
\author{Elaheh Fata, Igor Kadota and Ian Schneider}

\begin{document}
%\onecolumn

\maketitle

\begin{abstract}
Electricity is bought and sold in wholesale markets at prices that fluctuate significantly. Short-term forecasting of electricity prices is an important endeavor because it helps electric utilities control risk and because it influences competitive strategy for generators. As the ``smart grid'' grows, short-term price forecasts are becoming an important input to bidding and control algorithms for battery operators and demand response aggregators. While the statistics and machine learning literature offers many proposed methods for electricity price prediction, there is no consensus supporting a single best approach. We test two contrasting methods for predicting electricity prices--regression decision trees and recurrent neural networks (RNNs)--and compare them to a more traditional ARIMA implementation. We conduct the analysis on a challenging dataset of electricity prices from ERCOT, in Texas, where price fluctuation is especially high. We find that regression decision trees in particular achieves high performance compared to the other methods, suggesting that regression trees should be more carefully considered for electricity price forecasting.   
\end{abstract}


\section{INTRODUCTION}\label{sec.Intro}
\IEEEPARstart{T}{he} cost of supplying electricity varies constantly according to factors like demand, fuel prices, and the availability of power plants and renewable energy. Demand is a significant driver of the electricity price which itself depends on a variety of factors such as time of the day, day of the week, and the weather. As renewable energy, e.g. wind and solar energy, is increasingly incorporated into the electricity grid, it adds new uncertainty to electricity provision. For instance, if wind blows in a given hour then the short-term price of supplying electricity decreases.  

In recent years, electricity markets in different countries have been deregulated, allowing for the introduction of dynamic pricing. For most residential consumers in the United States, a local utility procures energy on their behalf through wholesale electric power markets with varying rates. However, in some markets, commercial and industrial customers are directly exposed to wholesale rates. 

Long-term forecasting is used by power utilities for risk management and investment profitability analysis, while short-term forecasting has been used by consumers and producers to derive bidding strategies. Much attention has been paid to new ``smart grid'' technologies that could improve grid flexibility, like batteries for energy storage or technologies to engage customers in demand response. % by shifting their demand in response to real and predicted prices. 
Operators of renewable energy generation might also use price prediction to determine their optimal energy offerings, given some uncertainty about the availability of energy in the near future \cite{Schneider2016} \cite{Schneider2017}.
The success of these technologies depends, in part, on their ability to predict short-term electricity prices and optimize their operation in order to profit from fluctuations in energy prices. For this reason, the interest in the development of price forecasting tools has increased \cite{FEIJOO201627,Zareipour,LIU2013152,WERON2008744,OBES632,KNITTEL2005791,ARIMA1,Szkuta,YAMIN2004571,CATALAO20071297,LIN20102707,CATALAO20111061,OSORIO2014363,Vardakas2015,WERON20141030}. \textbf{Because of its growing importance for energy management and the ``smart grid'', the focus of this paper is on short-term forecasting}.

\begin{figure}[hb!]
\begin{center}
\includegraphics[height=4cm]{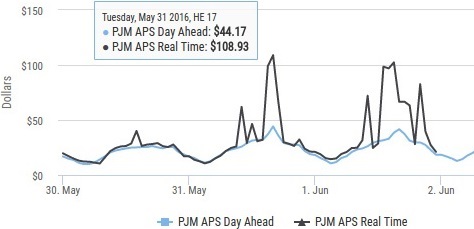}
\end{center}
\vspace{-1em}
\caption{Comparison of day-ahead and real-time prices from PJM in the period between 05/30/2016 and 06/02/2016. Figure from the \href{http://blog.kwantera.com/the-importance-of-day-ahead-real-time-energy-markets}{website}.}\label{fig.DA_vs_RT}
\end{figure}

For planning purposes, most system operators run two sequential electricity markets: day-ahead and real-time. The day-ahead energy market allows participants to secure the price of electricity one day in advance and hedge against real-time price fluctuations through a forward contract. The real-time market allows participants to buy and sell electricity throughout the operation day. Figure \ref{fig.DA_vs_RT} illustrates the evolution of the day-ahead and real-time prices over time in the PJM region, which covers much of the mid-Atlantic region in the United States.  

In most U.S. markets, the day-ahead market for electricity is cleared hourly, and the the real-time price for electricity varies every hour or every fifteen minutes. The real-time price can be significantly different from the day-ahead price due to unexpected fluctuations in the demand or because of a malfunctioning components of the grid (network outage). In general, electricity prices have a non-stationary mean and variance, which makes their prediction especially challenging \cite{FEIJOO201627}. \textbf{In this paper, we develop and compare algorithms that predict short-term real-time electricity prices} based on the available information, namely past prices, past demand, and current time/date information, wind predictions, and recent wind energy availability.%Specifically, we seek to predict real-time prices in the near future, using data from past and present events.

In the literature, three classes of techniques are used for predicting short-term electricity prices: classical statistical methods, data driven models and hybrid techniques. 
Classical statistical methods such as regression and Auto-Regressive Integrated Moving Average (ARIMA) \cite{Zareipour,LIU2013152,WERON2008744,OBES632,KNITTEL2005791,ARIMA1} are well-known and widely used to predict time-series data. The main benefit of such methods is their simplicity and possibility of further analysis due to their explicit prediction function.
Data Driven models such as Neural Networks learn the structure of the problem, namely the relationship between input and output, from data. Some papers that utilize this technique are \cite{Szkuta,YAMIN2004571,CATALAO20071297,LIN20102707}. Hybrid techniques combine both approaches \cite{FEIJOO201627,CATALAO20111061,OSORIO2014363}. Literature surveys can be found in \cite{Vardakas2015,WERON20141030}

This paper explores both classical statistical methods and data driven models. In particular, we develop prediction algorithms based on ARIMA, Regression Decision Trees and Neural Networks. We compare their performance in terms of prediction accuracy and then discuss their advantages and drawbacks with respect to feature engineering, computational complexity and suitability to the price prediction task.

The algorithms in this research are designed and evaluated based on datasets obtained from \href{http://www.ercot.com/about}{ERCOT}, the operator for the electricity system in Texas. The datasets contain time-stamped information over a two-year period (2014 and 2015), including real-time prices, real-time demand, demand forecast, day-ahead prices set by the market and wind conditions. The ERCOT dataset is particularly challenging for electricity price analysis for two reasons. First, while its data is publicly available it is not simple to manage or use. We elaborate on this in Section \ref{sec::data}. 

Second, the ERCOT data has a much higher maximum price than other markets in the United States and worldwide, and, partially as a result, it features very high price variance compared to other regions. For our dataset, the average price is $\$ 20.15$ with a standard deviation of $\$59.22$ and the maximum price is $\$5040$, 250 times the average price! In some parts of Texas, taking into account variation in demand, the top $2\%$ of most expensive hours make up $20\%$ of total annual electricity costs \cite{schneider_sunstein_2017}. Figure \ref{price_comparison} shows a comparison of mean, standard deviation, and max prices for three regions from 2014-2015. Price data from Denmark and from PJM (in the Mid-Atlantic United States) are often used for testing forecasting models, but they offer significantly lower standard deviation (especially proportional to their mean) and significantly lower peak prices compared to ERCOT. Thus, ERCOT represents a more challenging dataset for electricity price prediction that has not been tested previously in the literature.  

\begin{figure}
    \centering
\begin{tabular}{c|c|c|c}
	\centering
	Location & Denmark & PJM & ERCOT \\ 
	\hline
	Mean  & 26.94 & 38.53 & 20.15 \\
	SD &  13.08 & 44.71 & \textbf{59.22} \\
    Max & 117.37 & 1722 & \textbf{5040.40} \\
\end{tabular}
\caption{Mean, Standard Deviation (SD), and Max Prices (all in USD) for electricity prices at single price nodes in three separate locations.}
\label{price_comparison}
\end{figure}

We focus on prediction of a specific nodal price. Nodal prices are calculated at every major intersection of the transmission network in ERCOT, there are about 500 of them in ERCOT. Much existing research focuses instead on regional average prices, which display significantly lower variance since they are an average over multiple locations. Thus, the prediction of nodal prices poses a much more significant learning challenge. However, it is also much more relevant, since the nodal prices are ultimately what is used for the settlement of electricity consumption by generators at their particular location. 

The remainder of the paper is outlined as follows. In Sec.~\ref{sec.Prediction}, we formalize the objective of the price prediction task and describe the dataset in detail. In Sections \ref{sec.Time}, \ref{sec:trees} and \ref{sec.NN} we develop prediction algorithms based on ARIMA, Regression Decision Trees and Neural Networks, respectively, and discuss each method. The paper is concluded in Sec.\ref{sec.Conclusion}.

\section{PRICE PREDICTION MODEL}\label{sec.Prediction}
In this section, we describe the prediction model in detail. First, we present the loss function to be minimized, and then we describe the datasets that are utilized to train and validate the prediction algorithms.

\subsection{Accuracy Metrics}
Different metrics can be utilized to evaluate the performance of prediction algorithms. Let $P_t^\mathrm{real}$ and $P_t^{\mathrm{prediction}}$ be the actual and predicted electricity price at time $t$. Using this notation we describe some well-known accuracy metrics \cite{Vardakas2015}.

Two basic metrics are the Absolute Error and the Absolute Percentage Error. The Absolute Error at time $t$ is defined as $|P_t^\mathrm{real}-P_t^\mathrm{prediction}|$ and the Absolute Percentage Error is
\begin{equation}
\displaystyle\frac{|P_t^\mathrm{real}-P_t^\mathrm{prediction}|}{|P_t^{real}|} \; .
\end{equation}

Two metrics that consider the accuracy over a time-horizon with a total of $N$ predictions are the Mean Absolute Error and the Root Mean Squared Error. The Mean Absolute Error is given by
\begin{equation}\label{eq.MAE}
MAE = \frac{1}{N}\sum_{t=1}^{N}|P_t^\mathrm{real}-P_t^\mathrm{prediction}| \;,
\end{equation}
and the Root Mean Square Error metric is:
\begin{equation}\label{eq.RMSE}
RMSE = \displaystyle\sqrt{\frac{1}{N}\sum_{t=1}^N\left(P_t^\mathrm{real}-P_t^\mathrm{prediction}\right)^2} \; .
\end{equation}
MAE is widely used in the literature because of the extremely high variance of electricity prices. However, its derivative is non-continuous, which introduces additional challenges for implementation. We use both MAE and RMSE in this research.

\subsection{Dataset} \label{sec::data}
\begin{figure}[hb!]
\begin{center}
\includegraphics[height=5cm]{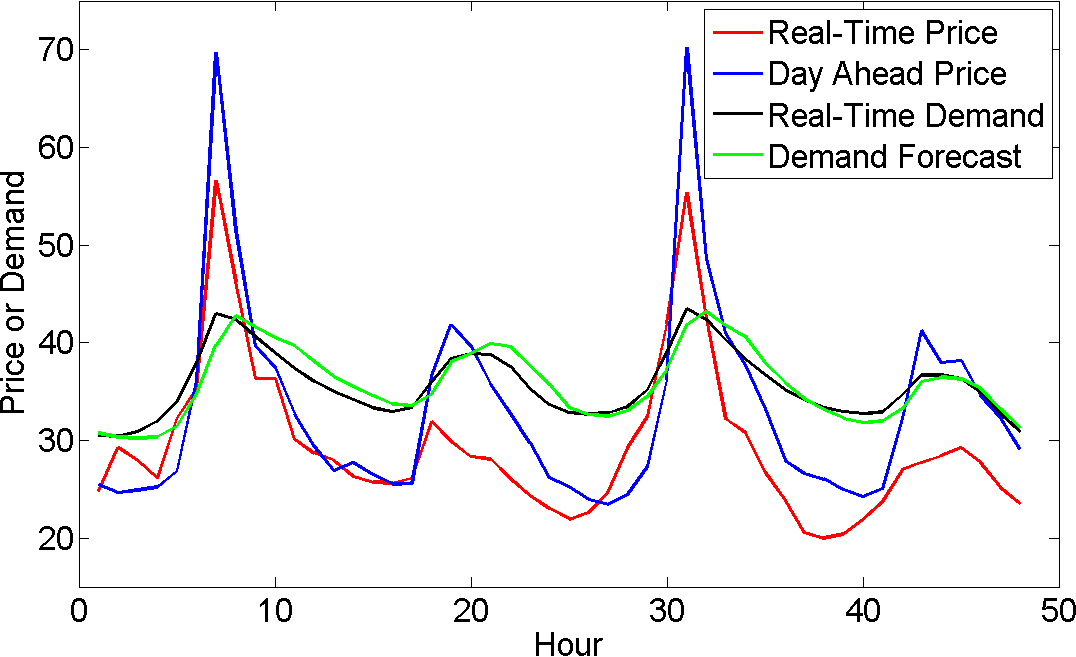}
\end{center}
\vspace{-1em}
\caption{Illustration of the ERCOT data with hourly information in the period between 01/15/2014 and 01/16/2014.}\label{fig.price_vs_demand}
\end{figure}

\begin{figure}[hb!]
\begin{center}
\includegraphics[height=5cm]{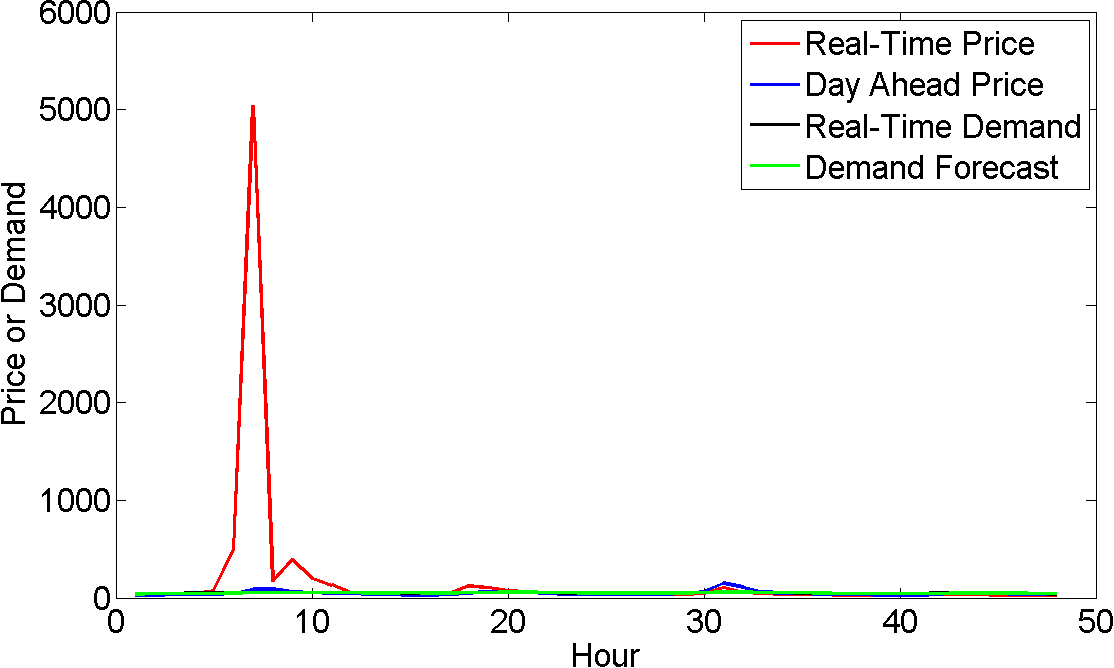}
\end{center}
\vspace{-1em}
\caption{Illustration of the ERCOT dataset with hourly information in the period between 01/06/2014 and 01/07/2014.}\label{fig.price_vs_demand_spike}
\end{figure}

Our dataset contains time-stamped information, including:
\begin{itemize}
    \item real-time prices in intervals of 15 minutes;
    \item real-time demand in intervals of 15 minutes;
    \item demand forecast in hourly intervals;
    \item day-ahead prices set by the market in hourly intervals;
    \item wind conditions and predictions in hourly intervals.
\end{itemize}
In Figures \ref{fig.price_vs_demand} and \ref{fig.price_vs_demand_spike} we compare the different information in the data over two representative periods\footnote{In Figures \ref{fig.price_vs_demand} and \ref{fig.price_vs_demand_spike} the demand data was re-scaled (divided by 1000) for the sake of clarity.}. 

The goal of the algorithms developed in this report is to predict real-time prices. To do so the algorithms use information on \emph{previous} prices $P_n^\mathrm{real}$, demand $D_n^\mathrm{real}$ and wind conditions $W_n$, for $n<t$; and information on \emph{previous and future} demand forecast $D_n^\mathrm{forecast}$ and day-ahead prices $DA_n$, for any time $n$. Algorithms should not rely on the entire historical data to predict price at time $t$. It is evident that the relevance of information from one hour is higher than the information from one month ago. For this reason, we define a past time-window $W_P$ and say that at time $t$ past information is available only in the interval $n \in \{t-W_P,\cdots,t-1\}$. Future information is only available up to a certain horizon. In particular, day-ahead prices are available only 24 hours in advance. Hence, we also define a future time-window $W_F$ and say that at time $t$ future information is available only in the interval $n \in \{t,\cdots,t+W_F\}$. 

Thus, the \textbf{features} utilized to obtain $P_t^\mathrm{prediction}$ are the real-time price, the real-time demand and the wind data in the interval $n \in \{t-W_P,\cdots,t-1\}$, in addition to the demand forecast and the day-ahead price in the interval $n \in \{t-W_P,\cdots,t+W_F\}$. Notice that $W_F$ and $W_P$ control the dimension of the input vector and, as a result, they influence running time of algorithms. One example of interest is $W_P=30$ hours and $W_F=4$ hours, for which algorithms can see information from the previous day and take advantage of some ``periodic'' events such as: at the beginning of business hours, the demand grows. Figure \ref{fig.price_vs_demand} shows that  $P_t^\mathrm{real}$ and $P_{t-24}^\mathrm{real}$ can be correlated. Moreover, Figure \ref{fig.price_vs_demand} demonstrates that the real-time price is correlated to the demand forecast and day-ahead prices. Hence, having $W_F=4$ should benefit the real-time price prediction. However, Figure \ref{fig.price_vs_demand_spike} illustrates a case in which data seems to be not correlated, thus \textbf{emphasizing the challenge of predicting short-term electricity prices}.

It is worth mentioning that the ERCOT dataset does not have complete data on every hour in 2014 and 2015 and some files have missing information. We chose to adjust the implementation of our data algorithms to ignore missing data, and in calculating our final loss values we also ignored periods with missing data (this always amounted to less than 1\% of total data). Due to the missing data, and some instances of repeated data, it was a significant challenge to clean and align the data, requiring hours of manpower and computational resources. This might explain why we haven't found any mention of price forecasting for the ERCOT data, even though it is feature-rich and publicly available. 

%\textbf{[Comments on cross-validation (?)]}

\section{ARIMA}\label{sec.Time}
In this section, we develop an ARIMA estimator based on the technique described in \cite{ARIMA1,ARIMA2}. The basic form of an ARIMA estimator is given by
\begin{equation}\label{eq.ARIMA}
    P_t^\mathrm{prediction}=\sum_{i=1}^k\alpha_iP_{t-i}^\mathrm{real}+\epsilon_t+\sum_{j=1}^q\theta_j\epsilon_{t-j} \; ,
\end{equation}
where $\alpha_i$ and $\theta_j$ are parameters associated with the Auto-Regressive and Moving Average parts, respectively, and $\epsilon_t$ are iid Gaussian random variables with zero mean and variance $\sigma^2$. Notice that ARIMA relies only on the past real-time prices in order to forecast $P_t^\mathrm{prediction}$. ARIMA is a very simple model and it does not use any of the exogenous features in our dataset.

To develop an ARIMA estimator, we follow a sequence of steps: 0) time-series data preprocessing, in particular, first order differencing; 1) model selection based on \eqref{eq.ARIMA}; 2) Maximum Likelihood estimation of the parameters $\alpha_i$, $\theta_j$ and $\sigma^2$ using the training set; 3) data forecast $P_t^\mathrm{prediction}$ using the trained ARIMA model on the validation set; 4) computation of the prediction error. The steps are repeated if the prediction error is unsatisfactory.

Recall form Section \ref{sec::data} that $P_t^\mathrm{real}$ is correlated with recent prices and also with the price at the same time in the previous day, $P_{t-24}^\mathrm{real}$. From this intuition, a suitable ARIMA model would be
\begin{align}\label{eq.ARIMA_chosen}
    P_t^\mathrm{prediction}=&\sum_{i=1}^2\alpha_iP_{t-i/4}^\mathrm{real}+\alpha_3P_{t-24}^{real}+\nonumber\\
    &+\epsilon_t+\sum_{j=1}^2\theta_j\epsilon_{t-j/4}+\theta_3\epsilon_{t-24}\; .
\end{align}
To predict $P_t^\mathrm{real}$, this model leverages information from the real-time prices in the past half-hour $\{t-1/4,t-1/2\}$ and in the previous day $\{t-24\}$.

\begin{figure}[hb!]
\begin{center}
\includegraphics[height=3cm]{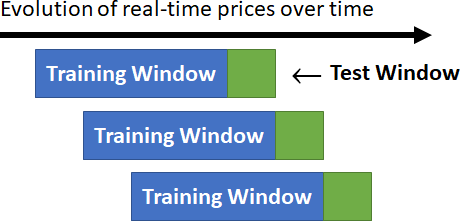}
\end{center}
\vspace{-1em}
\caption{Illustration of the cross-validation technique utilized for ARIMA.}\label{fig.Rolling}
\end{figure}

To estimate the parameters $\alpha_i$, $\theta_j$ and $\sigma^2$ and evaluate the accuracy of the trained ARIMA model, we use a cross-validation technique called \emph{forecast evaluation with a rolling origin} \cite{Rolling}, see Figure \ref{fig.Rolling}. The idea is to train the parameters based on the ARIMA model in \eqref{eq.ARIMA_chosen} and the values of $P_{t}^\mathrm{real}$ within the training window. Then, we use the trained ARIMA model to predict $P_t^\mathrm{prediction}$ in the test window and compute the error $P_t^\mathrm{real}-P_t^\mathrm{prediction}$. Next, we slide both the training window and the test window by the length of the test window and repeat the process. When the windows reach the end of the dataset, we can compute the overall MAE \eqref{eq.MAE} and RMSE \eqref{eq.RMSE} associated with the rolling Test Set. For the ERCOT dataset and the ARIMA model in \eqref{eq.ARIMA_chosen} with $Training Window=20$ days and $Test Window=5$ days, we obtain $MAE=5.09$ and $RMSE=23.39$.

\begin{figure}[hb!]
\begin{center}
\includegraphics[height=4.5cm]{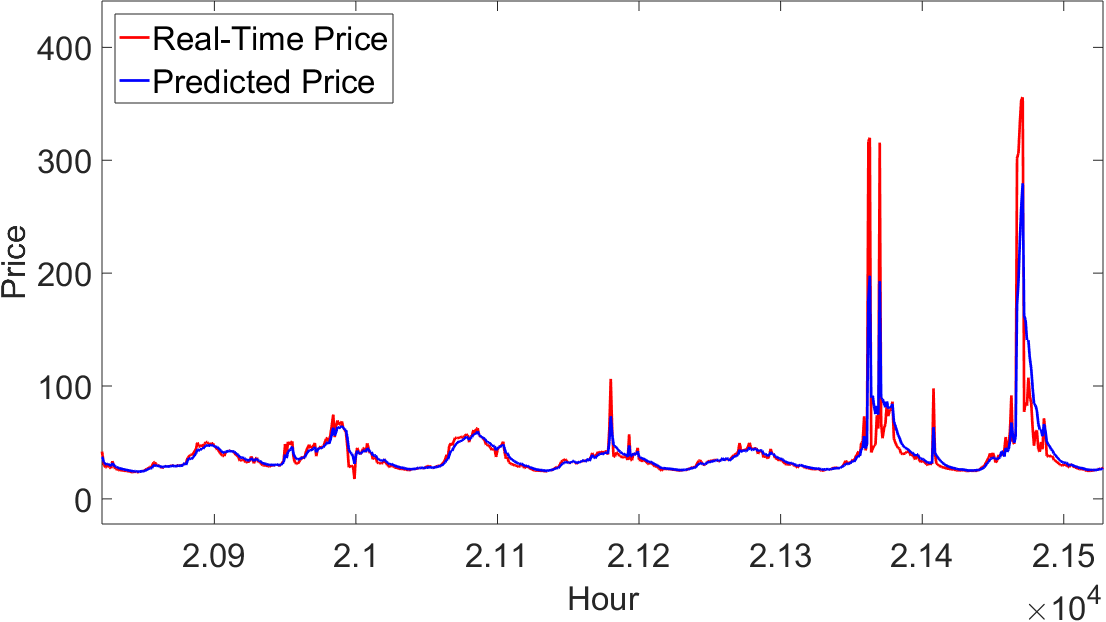}
\end{center}
\vspace{-1em}
\caption{Comparison of $P_t^\mathrm{real}$ and $P_t^\mathrm{prediction}$ over time.}\label{fig.ARIMA_forecast}
\end{figure}

Figure \ref{fig.ARIMA_forecast} displays the evolution of both $P_t^\mathrm{real}$ and $P_t^\mathrm{prediction}$ over time. In general, the ARIMA model accurately predicts the behavior of the real-time price. However, the prediction is noticeably less accurate when there are price peaks.
\begin{figure}[hb!]
\begin{center}
\includegraphics[height=4.5cm]{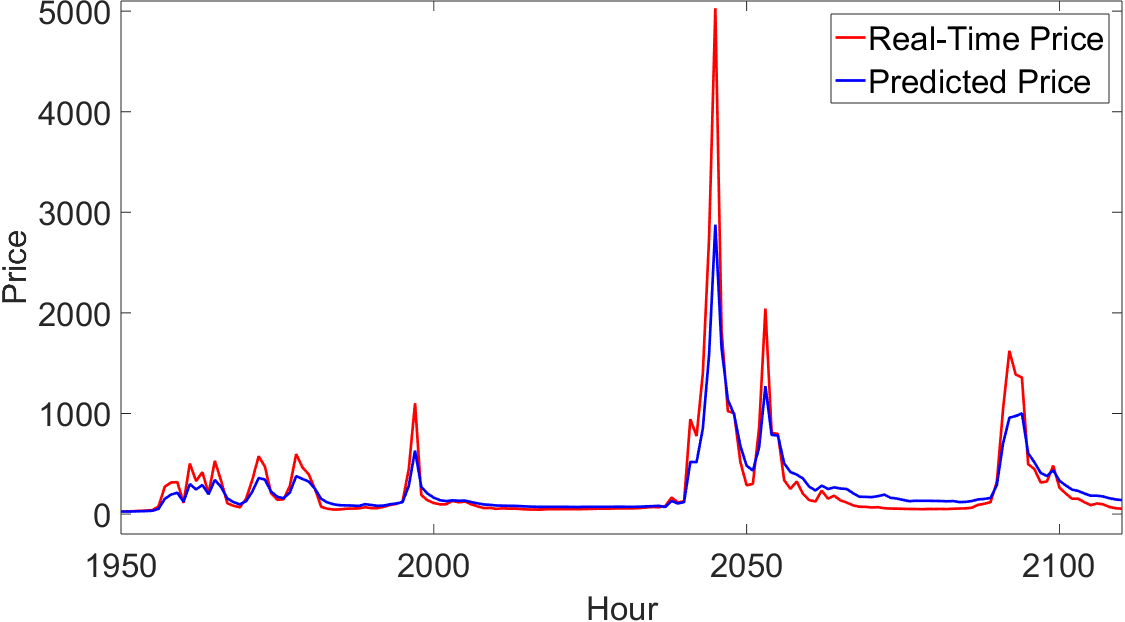}
\end{center}
\vspace{-1em}
\caption{Comparison of $P_t^\mathrm{real}$ and $P_t^\mathrm{prediction}$ over time. The peak in this picture is the same as the one in Fig.\ref{fig.price_vs_demand_spike}}\label{fig.ARIMA_forecast_peak}
\end{figure}

The main advantage of ARIMA is that it is based on a simple linear function \eqref{eq.ARIMA} that can be analyzed and can provide intuition into the prediction model. On the other hand, this simplicity is also its main disadvantage. In general, real-time prices are nonlinear in their features \cite{Vardakas2015} which cannot be represented by the linear ARIMA model. Hence, nonlinear models such as Regression Decision Trees and Neural Networks might achieve better accuracies. Next, we discuss those two nonlinear models in depth. 

\section{Regression Decision Trees}
\label{sec:trees}
We now turn our focus to different variations of regression trees and how they can be used for real-time price forecasting. 
The problem with naively implementing a regression model for electricity price prediction is that there are important underlying patterns in the data which cannot be explored by simple regression. 
%These underlying patterns are mainly due to the seasonality of the electricity price. 
For instance, electricity consumption during weekdays in July should have similar patterns, while the consumption during a weekend in November should have a different pattern, mainly due to the weather conditions and business intensity. %, or weekend consumption is different from weekday consumption. 
One way to explore those patterns is to utilize an unsupervised learning technique, such as decision trees, to separate the data into subclasses with similar electricity prices and then regress on each subclass. These methods can be seen in as similar to recent literature in electricity price prediction that combines SVM for classification with linear techniques for price prediction within each class \cite{zhang2011day}. However, the decision tree implementation has distinct advantages because it does not require the designer to specify the features that divide the data; instead it (ideally) chooses the features and classes that provide the most help for reducing the loss function. 
%determine clusters with similar electricity price patterns and then regress on each cluster. Another way to separate the data into subclasses is to partition the dataset using decision trees. 

{Binary regression decision trees}, {bagged regression trees} and {boosted regression trees} represent three widely used non-parametric approaches to predict demand and price~\cite{reston'17DT,yu'10DT,tso'07DT}. %These models are particularly useful as they are interpretable for managers of power companies. For simplicity of the model we assume that the electricity price at each node (i.e., each geographical point) is independent from the electricity prices at other nodes, hence we create a decision tree for each node independently. 
The features we use in our model to predict the real-time price at time $t$ are date/time information associated with $t$ (i.e. year, month, day of the year, day of the week, hour, etc.), day-ahead prices starting from the last eight hours to four hours ahead of the prediction time $t$ and real-time prices, wind, real-time demand and demand forecast information from the last eight hours. These particular time windows, namely $W_P=8$ hours and $W_F=4$ hours, were chosen by trying different time windows and observing the features that played important roles in the structure of decision trees.

\begin{figure}[h]
\begin{center}
\includegraphics[scale=0.2]{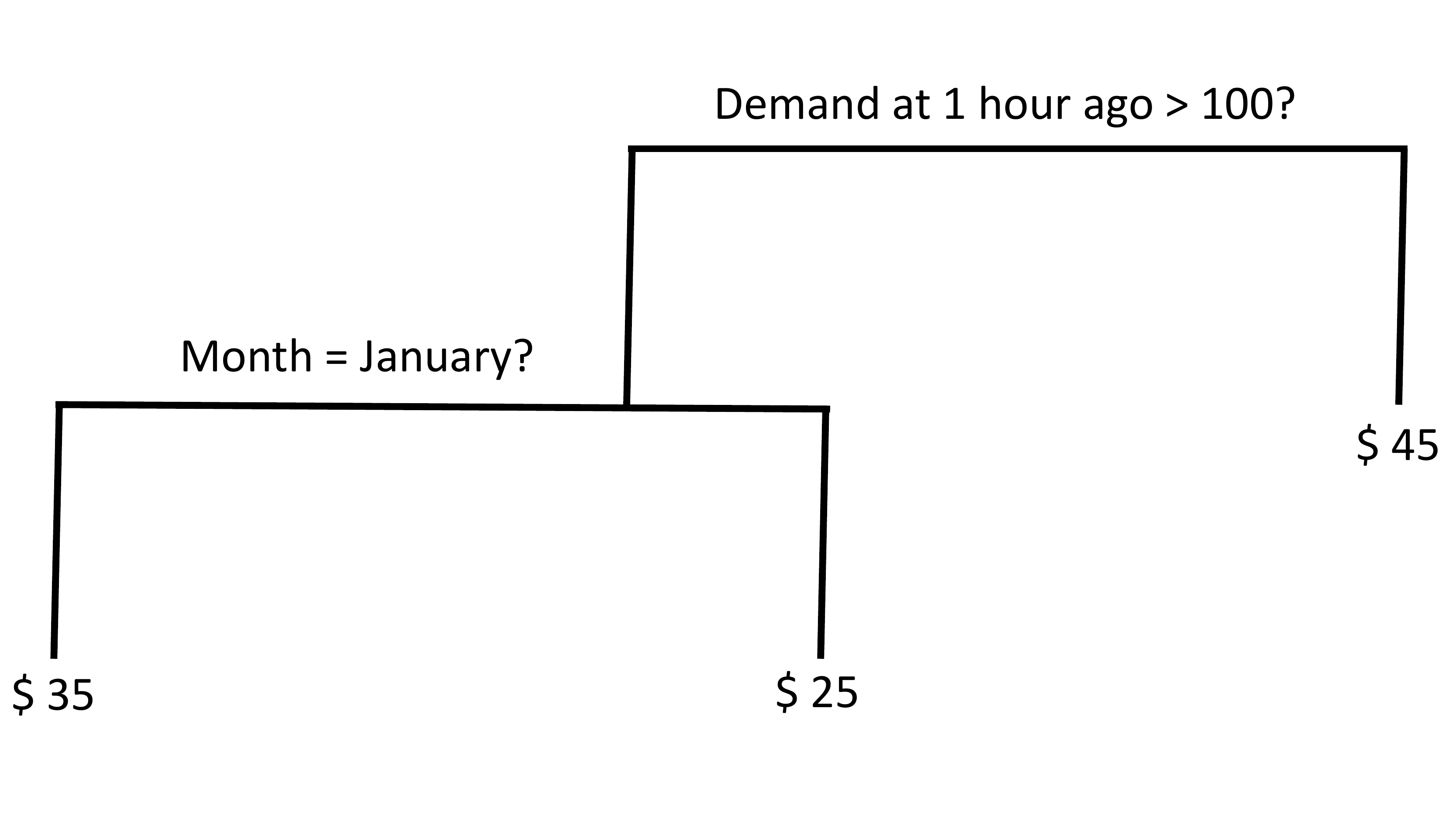}
\end{center}
\vspace{-1.5em}
\caption{Example of a binary decision tree.}
\label{fig:mytree}
\end{figure}

Next, we briefly introduce regression trees, bagged regression trees and boosted regression trees. We refer the reader to~\cite{bishop'00} for additional information. A simple binary decision tree for price prediction is illustrated in Figure~\ref{fig:mytree}. We start at the top of the tree and ask whether the electricity demand at the last hour was more then 100 units. If the answer is yes, we move to the right branch of the tree and predict the price of \$45 for the current time; otherwise, we move to the left branch and ask whether the prediction is done for the month of January. If the answer is yes, we predict the price of \$25, if the answer is no, our forecast is \$35. As seen in the example, one main benefit of decision trees is their ease of interpretation. %The simplicity of decision trees makes them interpret for those with little or no machine learning background, hence they are useful in big industries such as electricity generation.

A well-known problem associated with decision trees is over-fitting the training data by growing large and deep trees. We address this challenge in three different ways. In Section \ref{subsect:RDT}, we set bounds on model parameters such as minimum number of data elements in each leaf and maximum number of splits in the tree, in order to control the growth of the regression tree. The value of these bounds are found using cross-validation. In Section \ref{subsect:randforest}, we avoid over-fitting and reduce variance by implementing bagged regression trees. Bagging is an ensemble method in which the Training Set is divided into $b$ sets using \emph{sampling with replacement} and then a single regression tree is built for each of these $b$ sets. Once all these trees are built, the prediction of an unseen data point is given by the average of the predictions of each of the $b$ trees. This method is known to reduce prediction variance and to avoid over-fitting. Finally, in Section \ref{subsect:boostedtrees}, we implement boosted regression trees and compare the result with both previous approaches. Boosted regression trees are an ensemble of weak prediction models that are iteratively designed to make a single strong prediction model. 

\vspace{-1em}
\subsection{Single Regression Decision Trees}
\label{subsect:RDT}
In this section, we develop a single regression tree on the entire training set and optimize the tree's parameters using cross-validation. First, we randomly choose 70\% of the ERCOT dataset as the training set and the remaining as the test set. Then, we use $k$-fold cross-validation to choose the best parameters for the regression tree. In the $k$-fold cross-validation, the training set is divided into $k$ subsets. One of the $k$ subsets is selected as the validation set and training is performed on the remaining $k-1$ sets. This process is repeated for each of the $k$ subsets. The final validation result is obtained by averaging the $k$ validations.

\begin{figure}[h]
\begin{center}
\includegraphics[scale=0.3]{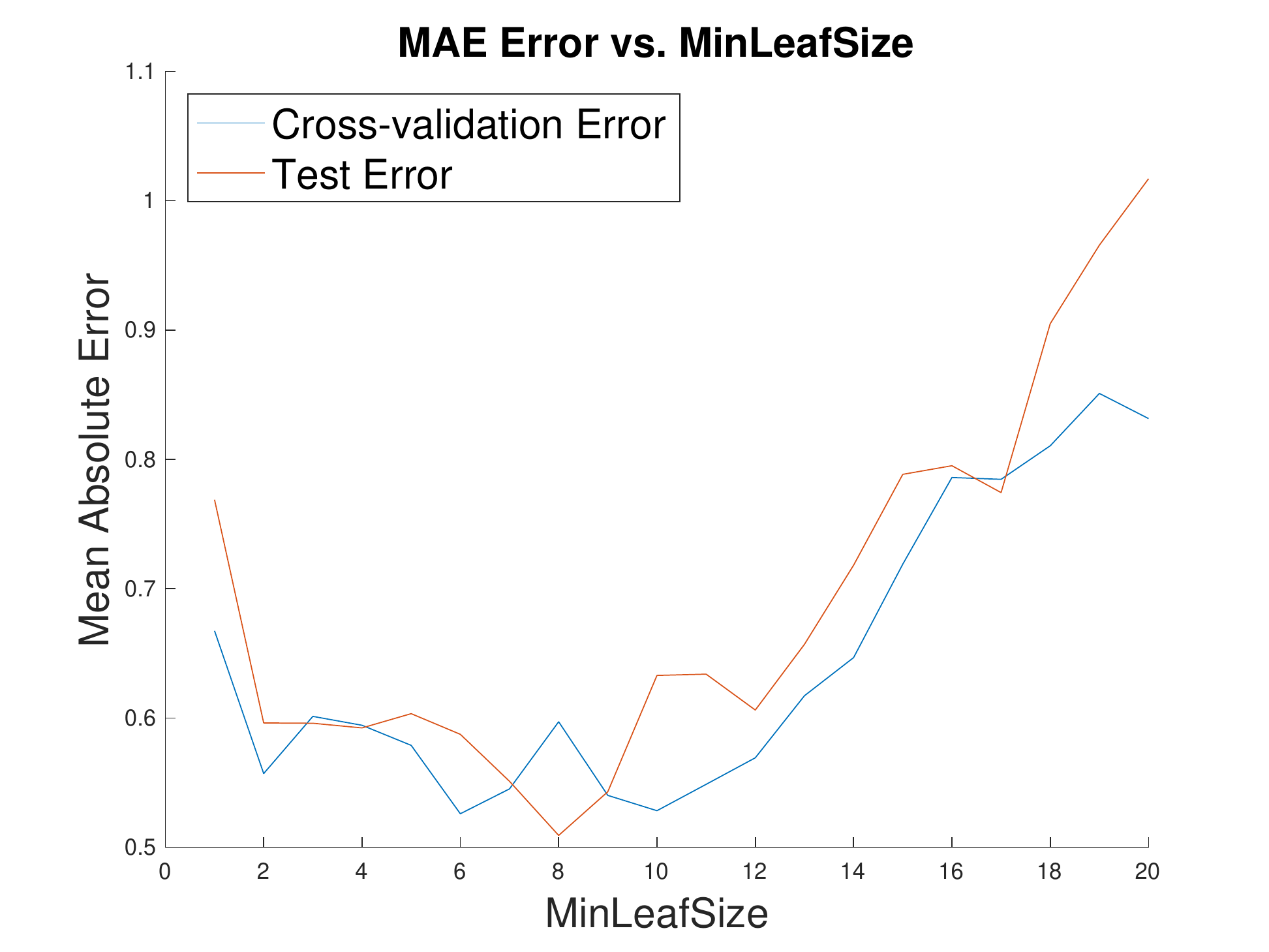}
\end{center}
\vspace{-1em}
\caption{Cross-validation and test MAEs for different bounds on min leaf size.}
\label{fig:rdt}
\end{figure}

To avoid over-fitting we restrict two parameters of the decision tree: minimum number of data elements in each leaf and maximum number of splits in the tree. Notice that if the maximum number of splits is not bounded, the tree can potentially split until there is only one data element in leaf. Our approach is to first train an unrestricted tree on the training set during the $10$-fold cross-validation process. We set the limit on the maximum number of splits as half the average number of splits in trained trees, i.e. $50$. We observe that limiting the number of splits increases the cross-validation mean square error by 5.6\%. For the minimum leaf size, we test different limits from 1 to 50 in the $10$-fold cross-validation and observe that a bound of $6$ provides good performance. Figure~\ref{fig:rdt} shows the impact of minimum leaf size on mean absolute error (MAE). Observe that minimum leaf size of $6$ provides the best cross-validation error. %certifying it is the best bound to use for the tree. 
Furthermore, despite the fact that we train only a single tree, this method achieves very accurate test errors (MAE less than \$1).

Lastly, we analyze the regression trees and verify which features were used more frequently in the decision process. Surprisingly, we observe that almost all branching was done on one feature: the price at two hours prior to the time of interest. Occasionally, other features such as time of the day and demand forecasts were used for branching but their role was not significant. This indicates that the two hours prior price feature is crucial in increasing information gain. %at each internal node which suggests that it is notably correlated to the desired price.

\subsection{Bagged Regression Trees}
\label{subsect:randforest}
\begin{figure}[h]
\begin{center}
\includegraphics[scale=0.28]{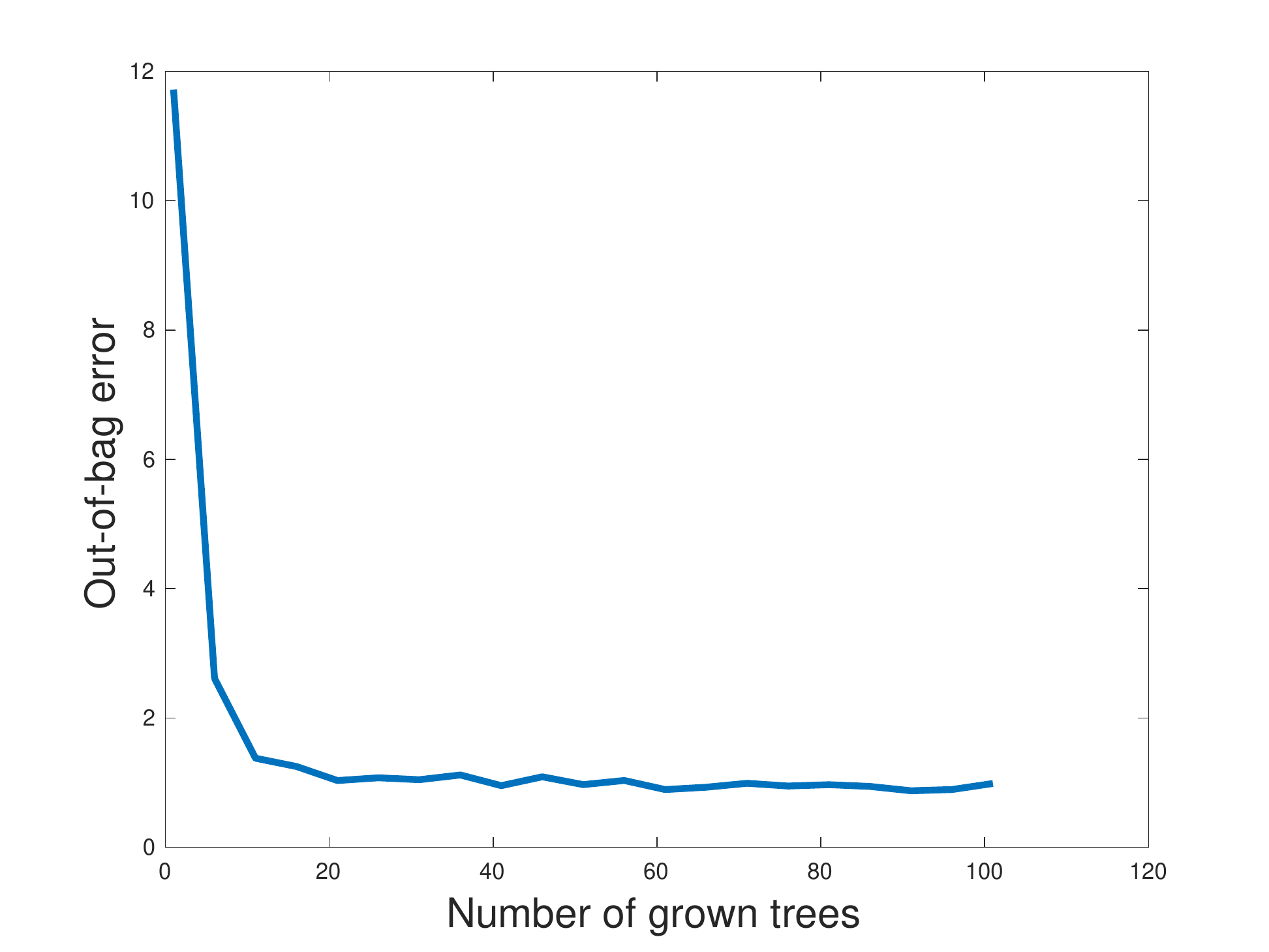}
\end{center}
\vspace{-1em}
\caption{Out-of-bag MAE for different number of trees.}
\label{fig:allOOB}
\end{figure}

Next we implement bagged regression trees, where data is divided into subsets and a regression tree is trained on each data subset independently; the output is the average of outputs of all these trees. Since bagging reduces variance to avoid over-fitting, we do not bound tree parameters and instead focus on finding the suitable number of trees to grow on the training set. For bagged regression trees it is conventional to use out-of-bag (OOB) analysis for cross-validation. Each tree in the bagged regression is trained on a subset of the training set, hence the remaining of the dataset is unknown to that tree and can be used as validation set to it. In OOB analysis, each training data point is tested on all the trees that did not have that point in their training set ~\cite{randomforest'01}. To find the optimal number of trees we find the mean absolute error of all training data points for bagged regression trees with size between 1 and 100 and pick the size associated to the smallest mean absolute OOB error, see Figure~\ref{fig:allOOB}. It can be seen that bagged regression trees of size 60 has a reasonable mean absolute OOB error and adding more trees to the bag does not improve the model much, hence we bag 60 trees in our regression. The mean absolute OOB and test errors associated to the bagged regression trees are \$0.89 and \$1.03, respectively. Observe that the test error for the bagged regression trees compared to a single regression tree in Section~\ref{subsect:RDT} is slightly larger; however, the difference is not substantial, and the bagged regression model could generalize better to new datasets. 

\begin{figure}[h]
\begin{center}
\includegraphics[scale=0.28]{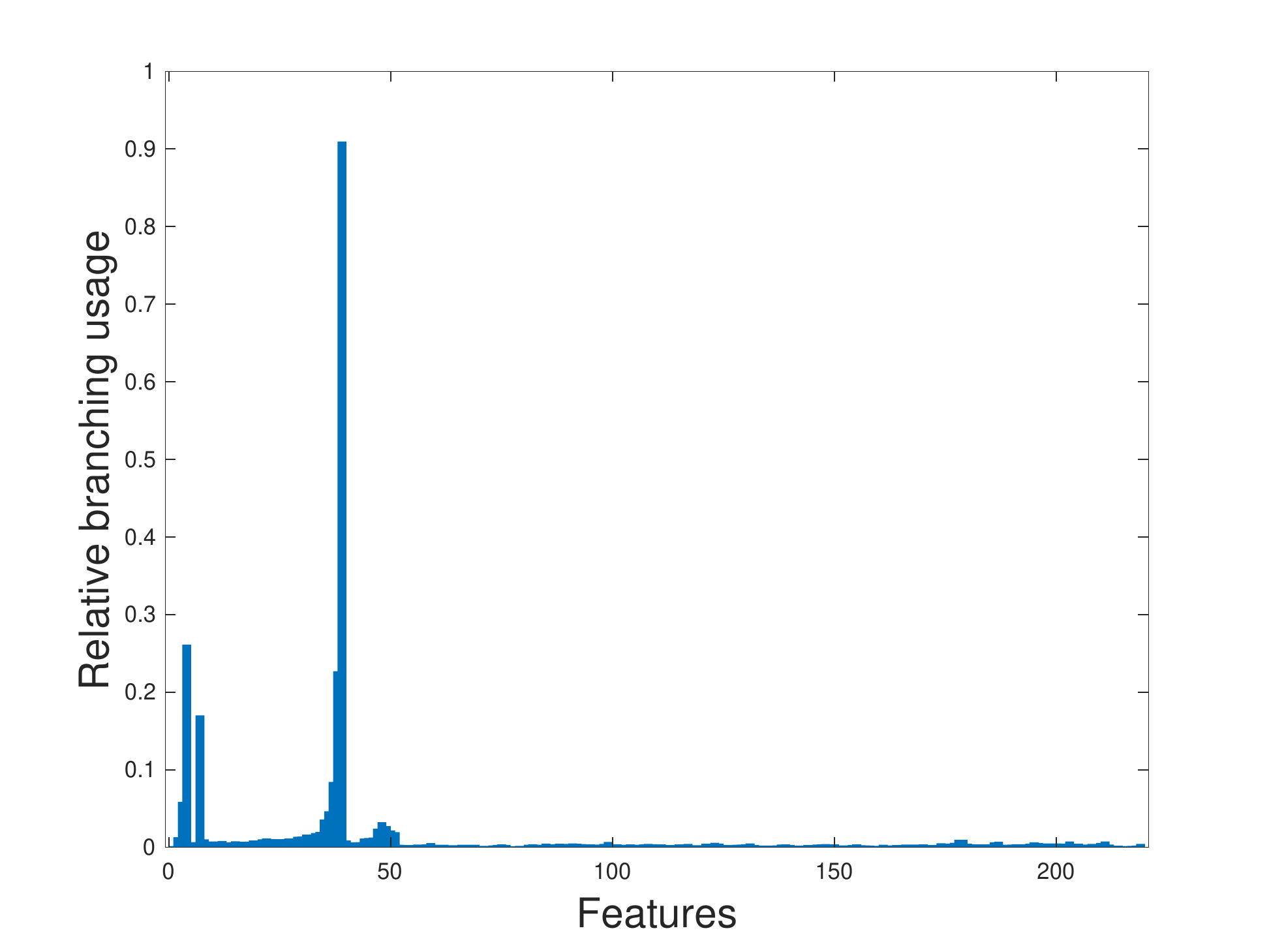}
\end{center}
\vspace{-1em}
\caption{Relative usage of feats. for branching in bagged regression trees with 60 trees.}
\label{fig:randforest-feats}
\end{figure}

Figure~\ref{fig:randforest-feats} shows the relative usage of features for branching in our bagged regression trees with 60 trees. It can be seen that one feature is used dominantly for branching and a few other features are used about one fifth as much and the rest of features are minimally used. The top five features in the order of significance are price at fifteen minutes ago, day of the year, price half an hour ago, time and price at forty five minutes ago. In comparison with a single regression tree, bagged regression trees use more features for branching, which suggests that it is a richer model and can be used when there are more complex features available in the data.

\subsection{Boosted Regression Trees}
\label{subsect:boostedtrees}

In this section, we discuss boosted regression trees. Recall that boosted regression trees are an ensemble of weak prediction models that are iteratively designed to make a single strong prediction model which minimizes a loss function. In particular, we use \emph{Least Square boosting} (LSboost), where the loss function is the squared error and the weak estimators are trees with bounded number of splits.

Let $x$ and $y$ denote the input feature and the correct prediction, respectively. At each stage $i$ of boosting, LSboost improves the previous imperfect model $F_{i-1}$ by adding a weak estimator $h$ to it, i.e. $F_i(x)=F_{i-1}(x)+h(x)$. The process is as follows: initially there is no prediction model and the best prediction is the mean values of the data points, i.e., $F_0=\overline{y}$. Then, at each subsequent stage, the goal is to find the estimator $h$ that provides the best prediction of $y-F_{i-1}(x)$. Notice that $F_i(x)=F_{i-1}(x)+h(x)=y$ or, equivalently, $h(x)=y-F_{i-1}(x)$. Moreover, observe that $y-F_{i-1}(x)$ is the gradient of least square error $\frac{1}{2}(F(x)-y)^2$ with respect to $F(x)$. Hence, LSboost improves the model by minimizing the square error~\cite{LSboost}.

A modification to LSboost that yields remarkable improvement is to utilize shrinkage or learning rate. This modification changes the model into $F_i(x)=F_{i-1}(x)+\nu h(x)$, where $0<\nu\le 1$. It is known that small learning rates improve the performance of the model dramatically at the cost of slower training~\cite{LSboost}. Next, we employ four different learning rates: 0.1, 0.25, 0.5 and 1.

As discussed earlier, the weak learners are trees with bounded number of splits. Since our weak learners are binary trees, the maximum number of splits in a tree is one less than the number of data points. Therefore, we run LSboost for different bounds on the maximum number of splits, namely between 1 and a quarter of the training data, to avoid substantial over-fitting. Moreover, to reduce over-fitting we perform 3-fold cross-validation on the data.

\begin{figure}[h]
\begin{center}
\includegraphics[scale=0.45]{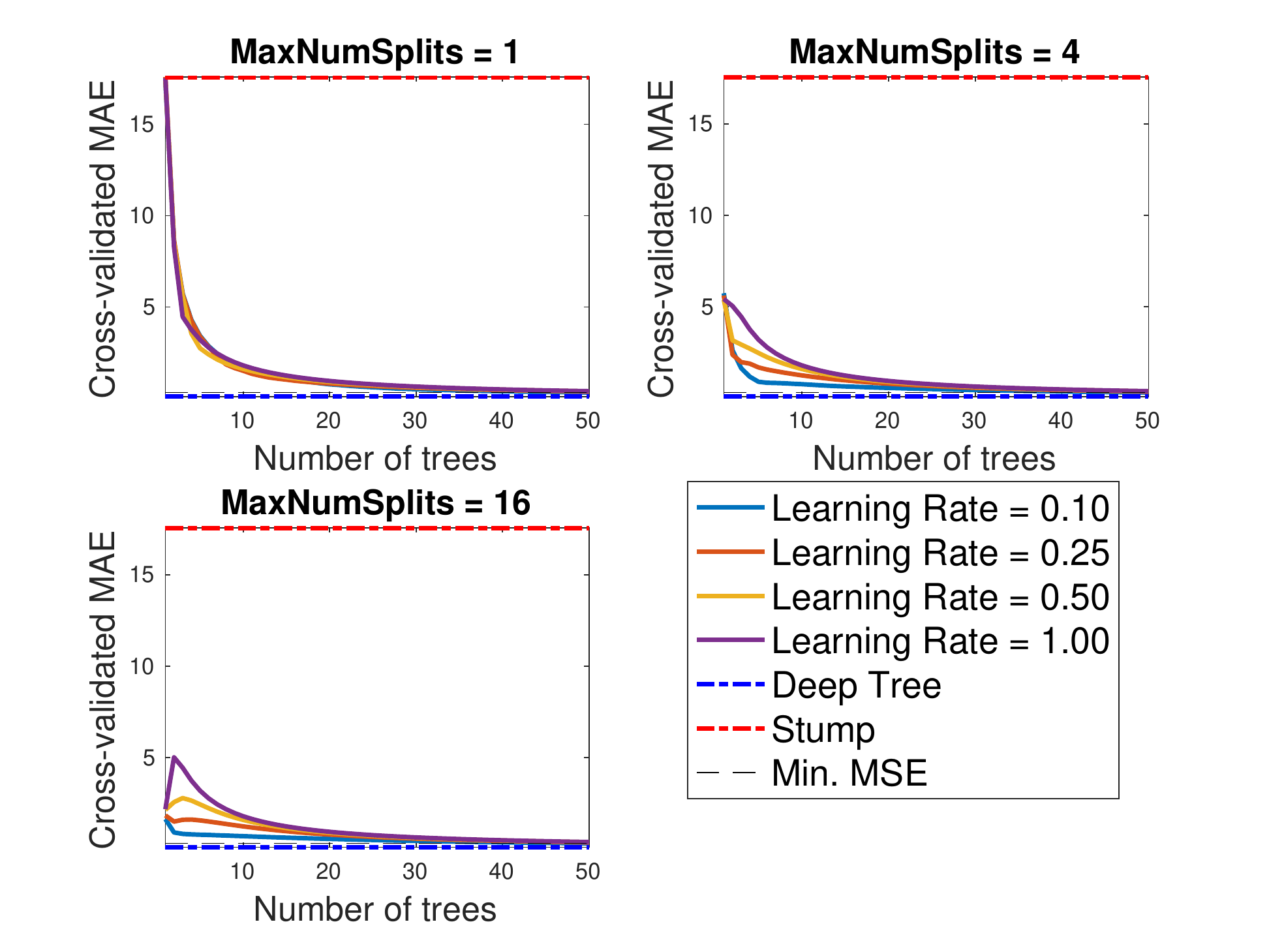}
\end{center}
\vspace{-1em}
\caption{Cross-validated MAEs for different learning rates and three different bounds on max number of splits of weak learners.}
\label{fig:boost-train}
\end{figure}

Figure~\ref{fig:boost-train} illustrates cross-validated mean absolute errors (MAE) for different number of boosting iteration using weak learners with maximum number of splits of 1, 4 and 16 and the four learning rates discussed above. Moreover, to better depict the performance of these boosted regression trees we plotted the MAE associated to deep trees (i.e., trees that exhaust the training data) and stumps (i.e., trees with only one splits). It can be seen that as the number of boosting iteration, i.e., number of trees, increases MAE improves. Moreover, using more sophisticated weak learners, that is, weak learners with higher bound on the number splits, results in smaller MAE, as expected. Deep trees will frequently over fit the data, while stumps do not have much power for accurate nonlinear prediction, as can be observed by their performance on our data. Lastly, we performed these boosted regression trees on the test data which provided similar results and accuracy.

\begin{figure}[h]
\begin{center}
\includegraphics[scale=0.3]{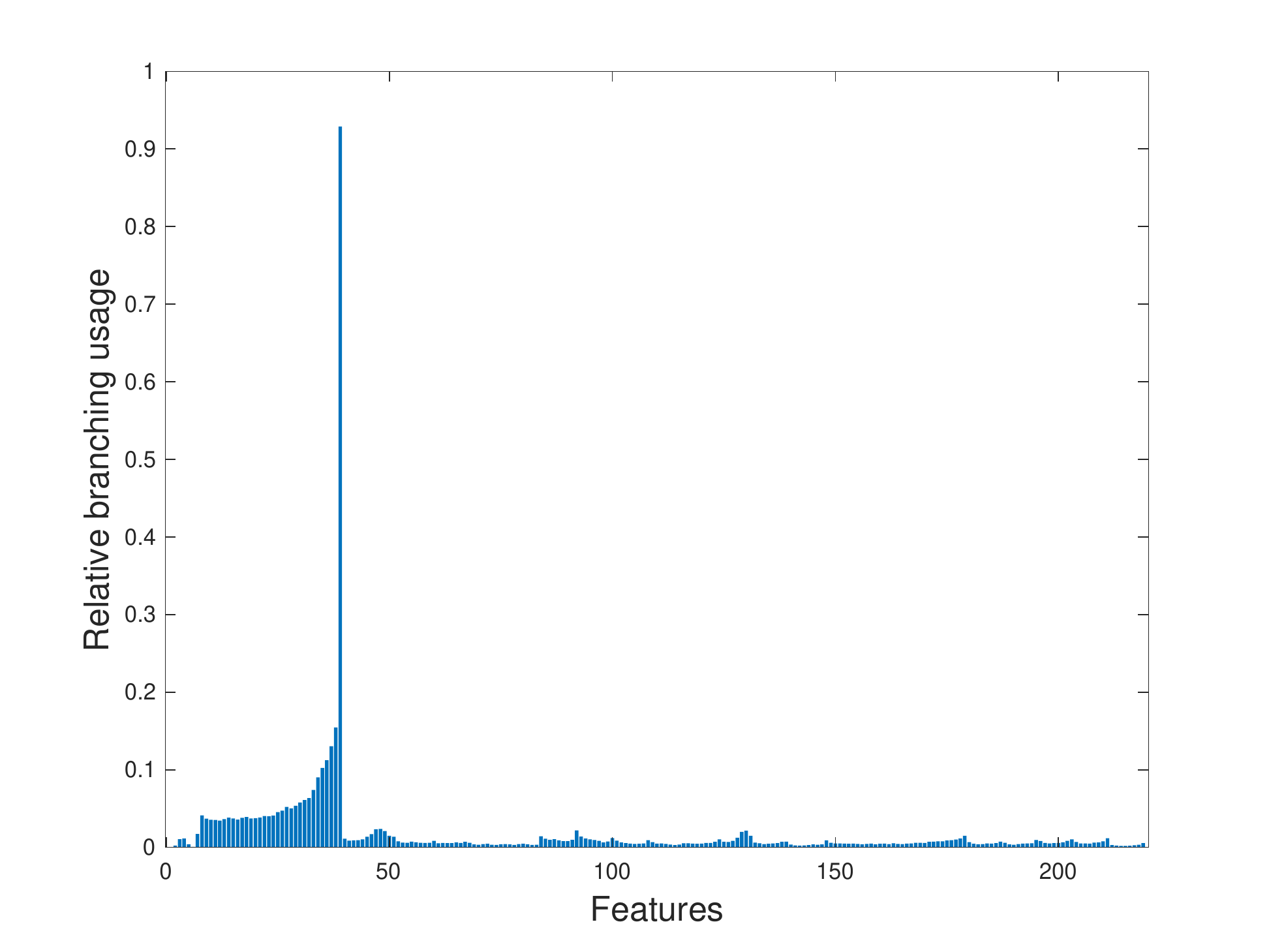}
\end{center}
\vspace{-1em}
\caption{Relative usage of feats. for LSboosting with 50 trees}
\label{fig:boost-feat}
\end{figure}

Moreover, Figure \ref{fig:boost-feat} depicts the relative usage of features for branching in LSboost algorithms with 50 trees. Again there is one dominant feature in that respect and other features are not used as much. The five top features in the order of usage are last fifteen minutes price, last half an hour price down to last seventy five minutes ago.

\noindent\textbf{Discussion:} 
All the three variation of regression trees we tried provided excellent accuracy. In particular boosted trees obtained higher accuracy which might be due their iterative improvement of the model. In particular boosted regression trees with numerous weak learners have low test errors and using them we achieved MAE of \$ 0.3 when 1024 weak learners were used for LSboosting. Moreover, it was interesting that all these methods used recent prices as their branching features and the remaining features did not play a significant role in the structure of trees.

\section{NEURAL NETWORKS}\label{sec.NN}
Variations of the Artificial Neural Networks (ANN) have been occasionally used in the literature for electricity price predictions due to their relatively high accuracy and capability of learning nonlinear relationships that are difficult to model with other  methods. While ``neural networks learns training data well, it may encounter large prediction errors in the test phase due to the time dependence of electricity prices '' \cite{FEIJOO201627}. We decided to design and implement a Recurrent Neural Networks (RNN), these have been occasionally but not frequently used in the literature \cite{WERON20141030}.

One of the main advantages of the RNN is that it tracks historical data, which allows it to automatically consider past data and predictions in its analysis. Thus it can provide nonlinear predictions related to both auto-regressive (AR) and moving average (MA) time-series predictions. By using a combination of historical data and exogenous inputs, the RNN is able to both consider current data and past prices when predicting current prices. However, in implementing the ARIMA and decision tree methods above, we use prepossessing to explicitly include previous values of certain features in the feature-list. %For instance, suppose there are $N$ features that might impact the electricity price at time $t$ and we also want to include the last $D$ periods of electricity prices and corresponding feature inputs. Then it is possible to create a feature vector of size $N D + D$ that includes the relevant past prices and features. This feature vector could be used for ARIMA, decision trees, or a more traditional ANN. 
There are two main disadvantages of RNNs for the purposes of short-term electricity price prediction. First, since it builds implicit functions, ``further analysis on the function forms such as sensitivity analysis is difficult'' \cite{LIU2013152}. This is true for all neural networks, not just RNNs. Second, RNNs suffer from the vanishing gradient problem, whereby it can be challenging to train their reaction to long-histories of data because the gradient will tend to vanish as you consider the effect of weights that operate on data farther back in the history.

\begin{figure}[h]
\begin{center}
\includegraphics[scale=0.25]{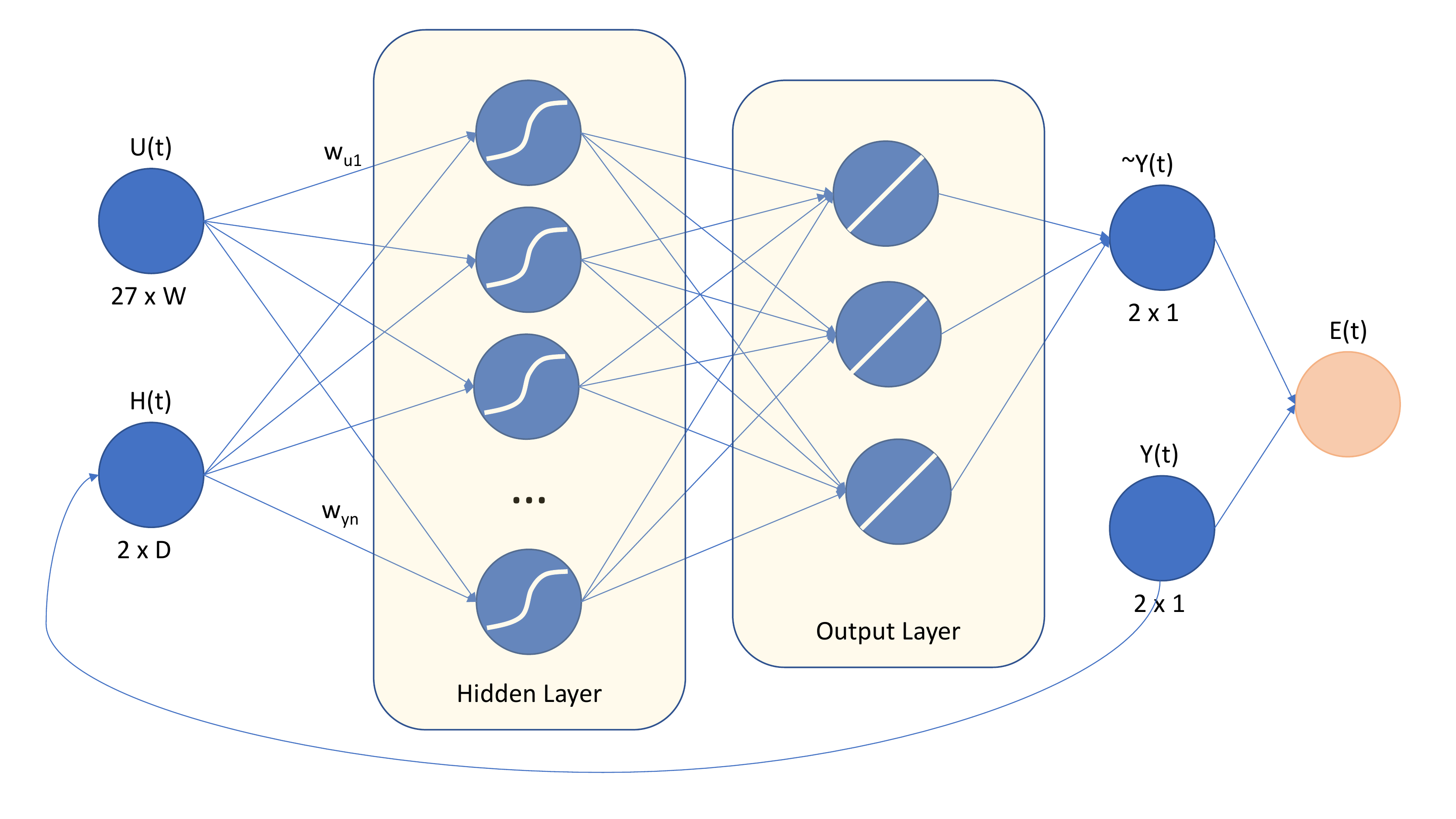}
\end{center}
\vspace{-1em}
\caption{Implemented RNN for electricity price prediction.}
\label{fig:RNN}
\end{figure}

\begin{figure*}
\begin{center}
\begin{tabular}{lllllll}
Loss Function & Training Method & Other Features & RSE\textsubscript{train} & RMSE\textsubscript{test} & MAE\textsubscript{train} & MAE\textsubscript{test} \\ \hline
MSE (inc. demand) & \begin{tabular}[c]{@{}l@{}}Levenberg-Marquardt\\ backpropagation\end{tabular} & - & 25.74 & 38.85 & 9.37 & 10.27 \\ \hline
MSE & \begin{tabular}[c]{@{}l@{}}Levenberg-Marquardt\\ backpropagation\end{tabular} & - & 31.38 & 41.87 & 8.14 & 8.80 \\ \hline
MAE & backprop & p = 10 & 67.63 & 58.79 & 16.67 & 16.96 \\ \hline
MAE & scaled conjugate gradient backprop & p = 50 & 53.13 & 78.79 & 6.78  & 7.42 \\ \hline
MAE & scaled conjugate gradient backprop & p = 100 & 45.97 & 74.64 & 5.39 & 6.20 \\ \hline
MAE & scaled conjugate gradient backprop & \begin{tabular}[c]{@{}l@{}}p = 200\\ M = 2\end{tabular} & 63.50 & 40.45 & 6.24  & 5.74 \\ \hline
MAE & scaled conjugate gradient backprop & \begin{tabular}[c]{@{}l@{}}p = 100\\ M = 3\end{tabular} & 50.45 & 57.81 & \textbf{4.99} & \textbf{5.38}    
\end{tabular}
\end{center}
\caption{Results for the RNN. Unless otherwise specified, the number of hidden layers $M=1$, and early stopping regularization are unused. For the first line, we also include the demand error in the loss function, and the RNN seeks to predict both price and demand simultaneously.}
\label{fig:RNN_results}
\end{figure*}

We implemented a multi-layer RNN with exogenous inputs, see Figure \ref{fig:RNN}. We started with the base implementation offered by \href{https://www.mathworks.com/help/nnet/modeling-and-prediction-with-narx-and-time-delay-networks.html}{MATLAB} and created and adjusted our code to focus the implementation for short-term price forecasting, including by the use of MAE. The exogenous input $U(t)$ at each time step contains $27$ features, which includes the current time and date, day-ahead prices, demand forecast information, and wind forecast information, i.e, all relevant data except for real-time price and real-time demand. The variable $W$ refers to the width (in terms of time) of exogenous inputs to include, so the prediction for price at time $t$ would use the exogenous inputs within times $(t-W,...,t)$. Unless otherwise noted, $W$ is fixed as 16. The output $Y(t)$ contains real-time prices.  %Despite our foremost interest in price, it is reasonable to organize the outputs this way as it makes it clear that the demand and price at period $t$ are realized simultaneously (since these are realized in the same market clearing). Thus, neither of them is available for prediction in period $t$, but they are available for predictions in periods $t+1$ and beyond.
Note that $Y(t)$ represents a $1\times (N+1)$ vector of the prices at $N$ locations and system demand at time $t$. Typically, $N$ is set to one, when are only trying to predict the price at a single location. The history $H(t)$ contains the prices and actual demand for the previous $D$ periods, unless otherwise specified, $D = 16$. The history is updated such that $H(t+1)$ is the concatenation of $Y(t+1)$ and all but the oldest value of $H(t)$.

In our implementation, the hidden nodes use the Sigmoid activation functions. Unless otherwise noted, there are 10 hidden units per layer, and a single hidden layer. We experimented with multiple hidden layers, and with varying numbers of hidden units per layer and did not see significant improvement. The output layer features one node with linear activation functions. Moreover, we use an early stopping technique based on the validation error. For a fixed $p$, if validation error increases for $p$ successive time steps we stop the training process. This can serve as a form of regularization that improves generalization and performance for insufficiently regularized implementations.

In short, the RNN predictor described above is given by:
\begin{equation}
    \hat{Y}(t) = f(U(t-W),...U(t), H(t);w)
\end{equation}
where $f(x;w)$ is the function applied by the RNN as a result of the network structure described in Figure \ref{fig:RNN} for weights $w$. Thus the natural loss function of mean absolute error is
\begin{equation}
    l(w) =  \sum\limits_{t = 1}^{T} |\hat{Y}_i(t) - Y_i(t)|.
\end{equation}
%Initially, we predicted the price and demand simultaneously. The benefit of this approach is that they can help improve generalization because they require the network to focus on two tasks simultaneously. Moreover, they are likely to use similar features for computation, so inputs to later nodes can be useful for predicting each of them. However, in our experience, our RNN obtained better results for predicting price when we adjusted the loss function so that it was only based on the price prediction error. Our intuition is that our efforts to improve generalization were already successful, so efforts to predict additional outputs just required the use of the resources of the network which limited its success for price prediction. 

The results of our RNN are described in Figure \ref{fig:RNN_results}. We initially utilized backpropagation to learn network parameters. This worked fine for our initial testing, but it proved to be very slow when early stopping was added to the model. %Learning time increased from around 45 minutes to over 90 minutes.  
As such, we utilized MATLAB's implementation of scaled conjugate gradient backprop \cite{Møller93ascaled} in order to speed up the results. This produced some of our best results when combined with three hidden layers and fairly relaxed early stopping criteria. Moeover, our early stopping technique successfully improved generalization. 

In our best implementation (both in terms of training and test accuracy for the MAE loss function), the test set loss was only about $7\%$ higher than the training set loss on the training. We want to add that we would have preferred to use deeper nets with less stringent stopping regularization, given that our test set loss did not start to diverge from the training set loss for any of our tested methods. However, in trying to expand to a deeper network, we encountered problems with insufficient memory, and reducing the early stopping criteria resulted in high increase in the running time without noticeable improvement. One effort to get around these challenges was to utilize a more efficient backpropogation algorithm, called Levenberg-Marquardt backpropogation \cite{hagan1994training}. Unfortunately, this algorithm turned out to be of limited use for us, because its MATLAB implementation does not allow for the use of loss functions besides MSE. When training with this algorithm, we successfully reduced the MSE of the training set well below what was achieved with other algorithms, see Figure~\ref{fig:RNN_results}. However, we also saw evidence of over-fitting as the test set error was 33\% higher than training set error. Most importantly, as Levenberg-Marquardt algorithm does not optimize MAE, its mean absolute errors were significantly higher for both the training and test sets compared to our best implementations.

\begin{figure}[h]
\begin{center}
\includegraphics[scale=0.32]{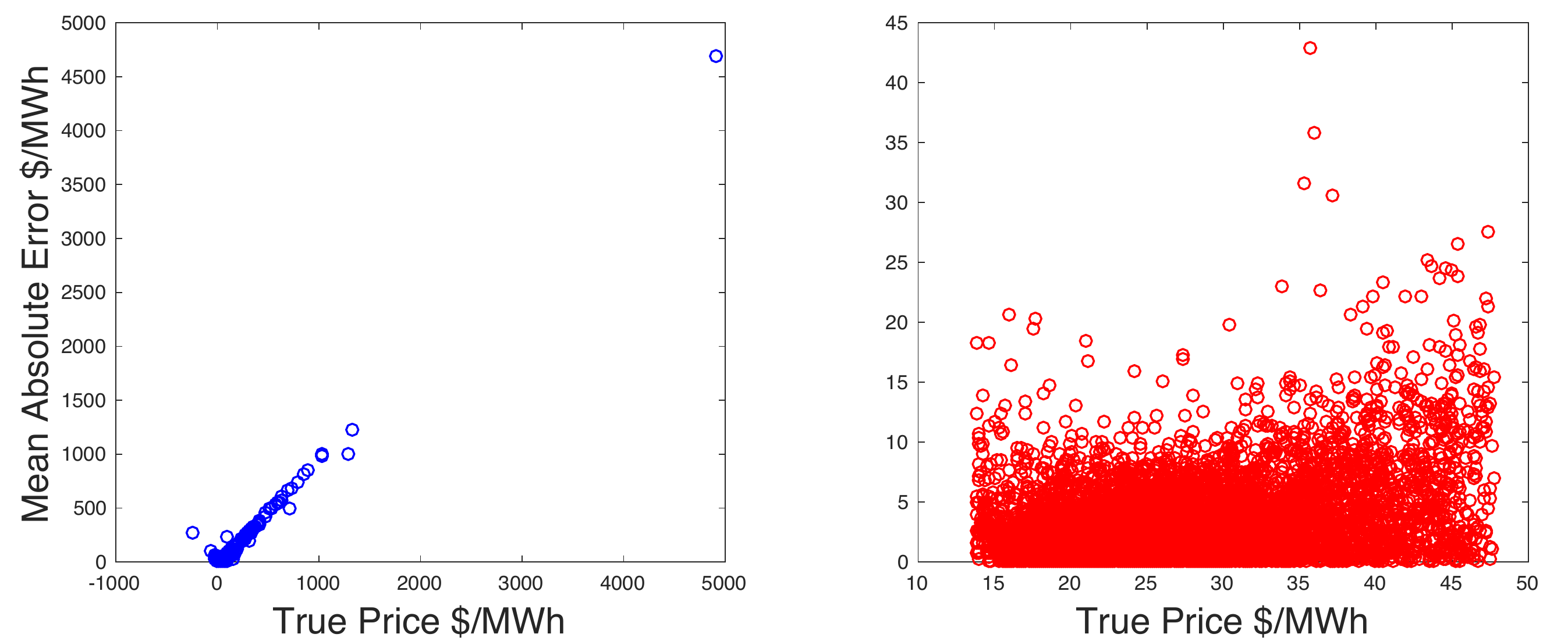}
\end{center}
\vspace{-1em}
\caption{The figure on the left shows true price versus mean absolute error for the test set for the best performing RNN implementation. The figure on the right excludes the 10\% of data points that were the farthest outliers.}
\label{fig:RNN_errors}
\end{figure}

When analyzing how our model learns, we find that it learns well for most prices but struggles to predict price spikes in the network. Figure \ref{fig:RNN_errors} shows the true price versus the mean absolute error for the best performing RNN implementation; we are interested in learning whether the RNN performs especially poorly when the true price is very low or very high. The graph on the left shows all data points, while the graph on the right excludes the 10\% of the data points that were the farthest outliers. A predictor that performs equally well on all data points will have similar errors regardless of the underlying value.  Taken together, these graphs suggest that the predictor accurately handles normal variation for most of the data points, but is unable to recognize or predict extremely high price periods.

%For the graph on the right, the predictions are not correlated to the true price. However, the graph on the left of Figure \ref{fig:RNN_errors} shows that as the price increases well above the mean, the increase in error is approximately linear with a coefficient of nearly 1! Essentially, the RNN is completely failing to predict the periods with extremely high price spikes; it predicts nearly average prices even in periods when the price spikes to $100$ times its average value. The failure of the RNN to accurately predict high-price spike periods contributes about $\frac{3}{4}$ of the total average absolute loss. 

Overall, as is typical with neural networks, we found that training parameters and initialization had important effects on performance. While the results for neural networks were not as impressive as we achieved using variations of regression decision tree models, this method shows promise for high-accuracy predictions, and could be improved further by continued adjustments and improvements to the implementation.

\section{CONCLUDING REMARKS}\label{sec.Conclusion}
We implemented three different methods for electricity price prediction, ARIMA, regression decision trees, and recurrent neural networks. We trained and tested our implementations on a challenging set of electricity price data from ERCOT, Texas, that featured high variance data as compared to electricity prices in other areas. Moreover, there were some missing and repeated data points in this dataset. We were able to develop RNN implementations that performed about as well as the standard ARIMA model, which is very popular for electricity price prediction. However, the nonlinear time-series model offered by the RNN did not substantially improve performance in our implementations, serving as a reminder that tuning and training neural networks can prove to be difficult or fruitless. On the other hand, the regression tree models performed extremely well, predicting typical prices and outliers with higher success than most comparative models. They achieved a 90\% reduction in test set error compared to the traditional ARIMA implementation, and they are comparable with the best methods in the electricity forecasting literature. While a direct comparison can not be made across datasets, our results show that regression tree implementations show high promise for accurate electricity price forecasting. 

\section*{ACKNOWLEDGMENT} Igor Kadota led the ARIMA implementation, Elaheh Fata led the regression tree implementation, and Ian Schneider led the RNN implementation. MATLAB was used for all implementations. Thank you to Hoon Cho for his help and comments on the research. 

\ifCLASSOPTIONcaptionsoff
  \newpage
\fi

\bibliographystyle{IEEEtran}
\bibliography{IEEEabrv}

\end{document}